\documentclass [11pt]{article}
\usepackage{amsmath,amsthm,amsfonts,amscd,eucal,latexsym,amssymb}
\usepackage{epsfig}   
\def\spa{\hskip -3pt}

\newsymbol\bt 1202           

\def\cI{{\ca I}}

\def\cG{{\ca G}}

\def\cH{{\ca H}}
\def\cD{{\ca D}}

\def\cS{{\ca S}}
\def\cA{{\ca A}}

\def\cX{{\ca X}}
\def\cW{{\ca W}}
\def\cK{{\ca K}}

\def\bC{{\mathbb C}}           
 
\def\bI{{\mathbb I}}

\def\bM{{\mathbb M}} 
 
\def\bR{{\mathbb R}} 

 \def\bF{{\mathbb F}}
\def\bS{{\mathbb S}}

\newsymbol\rest 1316         
 
\def\gH{{\mathfrak H}}

\def\beq{\begin{eqnarray}}
\def\eeq{\end{eqnarray}}

\newcommand{\ca}[1]{{\cal #1}}         

\newcounter{proposition}[section]
\newcounter{theorem}[section]
\newcounter{lemma}[section]
\newcounter{definition}[section]
\def\theproposition{\thesection.\arabic{proposition}}
\def\thetheorem{\thesection.\arabic{theorem}}
\def\thelemma{\thesection.\arabic{lemma}}
\def\thedefinition{\thesection.\arabic{definition}}

\def\s #1 {\section{#1}}

\def\ssa #1 {\ifhmode{\par}\fi\refstepcounter{subsection}
  \noindent {\bf\thesubsection}. {\em #1}.\quad
  \addcontentsline{toc}{subsection}{\protect\numberline{\thesubsection} #1}%
  }

\def\ssb #1 {\ifhmode{\par}\fi\refstepcounter{subsection}
  \noindent {\bf\thesubsection.} {\bf #1.}\quad
  \addcontentsline{toc}{subsection}{\protect\numberline{\thesubsection} #1}%
  }

\def\proposizione {\ifhmode{\par}\fi\refstepcounter{proposition}
  \noindent {\bf Proposition \theproposition}. \quad}
\def\teorema {\ifhmode{\par}\fi\refstepcounter{theorem}
  \noindent {\bf Theorem \thetheorem}. \quad}
\def\lemma {\ifhmode{\par}\fi\refstepcounter{lemma}
  \noindent {\bf Lemma \thelemma}. \quad}
\def\definizione {\ifhmode{\par}\fi\refstepcounter{definition}
  \noindent {\bf Definition \thedefinition}. \quad}

\begin{document} 
 
\hfill{\sl Preprint  UTM  686 - October  2005} 
\par 
\bigskip 
\par 
\rm

 
\par 
\bigskip 
\LARGE 
\noindent 
{\bf Bose-Einstein Condensate and Spontaneous Breaking of  
Conformal Symmetry
on Killing Horizons II} 
\bigskip 
\par 
\rm 
\normalsize 
 
 
\noindent {\bf Valter Moretti}  \\
\small
Dipartimento di Matematica dell'Universit\`a di Trento 
\& Istituto Nazionale di Alta Matematica ``F.Severi'',  Unit\`a Locale di Trento
\& Istituto Nazionale di Fisica Nucleare,  Gruppo Collegato di Trento,
via Sommarive 14, I-38050 Povo (TN), Italy\\
  E-mail: moretti@science.unitn.it\\ 
\normalsize 

 
\par 
\bigskip 

\noindent
\small 
{\bf Abstract}.  
 In a previous paper  local scalar QFT (in Weyl algebraic approach) 
  has been constructed on degenerate semi-Riemannian manifolds $\bS^1\times \Sigma$
 corresponding to the extension of Killing horizons by adding points at infinity to
 the null geodesic forming the horizon. It has been proved that the theory admits
 a natural representation of $PSL(2,\bR)$ in terms of $*$-automorphisms and  this representation
 is unitarily implementable if referring to a certain invariant state $\lambda$.
 Among other results it has been proved that the theory 
 admits a class of inequivalent algebraic (coherent) states
 $\{\lambda_\zeta\}$, with $\zeta\in L^2(\Sigma)$, which break part of the symmetry,
 in the sense that each of them is not invariant under the full group $PSL(2,\bR)$ and so
there is no unitary representation of whole group $PSL(2,\bR)$ which leaves fixed the cyclic GNS vector.
 These states, if restricted to suitable portions of $\bM$ are invariant 
and extremal KMS 
states with respect to a surviving one-parameter 
group symmetry.
In this paper we clarify the nature of symmetry breakdown. We show that,
in fact, {\em spontaneous} symmetry breaking occurs in the natural
sense of algebraic quantum field theory: if $\zeta \neq 0$, there is no unitary representation of whole 
group $PSL(2,\bR)$ which implements the $*$-automorphism representation of $PSL(2,\bR)$ itself 
in the GNS representation of $\lambda_\zeta$ (leaving fixed or not the state). 
\normalsize

\section{Summary of some achieved results.}

 In  \cite{MP6} local scalar QFT (in Weyl algebraic approach) 
 is constructed on degenerate semi-Riemannian manifolds $\bM = \bS^1\times \Sigma$
 corresponding to the extension of future Killing horizons by adding points at infinity to
 the null geodesic forming the horizon. Above the transverse manifold $\Sigma$
 has a Riemannian metric inducing the volume form $\omega_\Sigma$, whereas
 $\bS^1$ is equipped with the null metric.
 To go on, fix a standard frame (see section II A of \cite{MP6}) 
 $\theta \in (-\pi,\pi]$ on $\bS^1$ of $\bM = \bS^1\times \Sigma$ -- so that $\bS^1$ is realized as $(-\pi,\pi]$
 with the endpoints identified --
   and consider 
 the $C^*$-algebra of Weyl $\cW(\bM)$ generated by non-vanishing elements
 $V(\omega)$ in Eq. (12) in section II E in \cite{MP6}, the smooth forms $\omega$
 of the space $\cD(\bM)$
 being the space of  forms $\epsilon_\psi$ defined in Eq.(1) in section II A in \cite{MP6}.
 We shall exploit the group of $^*$-automorphisms  
 $\alpha$  defined in Eq. (19)
 in \cite{MP6} 
 representing the M\"obius group $PSL(2,\bR)$ viewed as a subgroup 
 of diffeomorphisms of $\bS^1$ and thus of $\bM$ (see Section III A of \cite{MP6}):
 \beq
 \alpha_g\left(V(\omega)\right) = V(\omega^{(g^{-1})})\label{0}\:, 
 \eeq
 $\omega^{(g)} := g^* \omega$ being the natural pullback action of $g\in PSL(2,\bR)$ on forms.
 In the following $\{\alpha^{(\cX)}_t\}_{t\in \bR}$ indicates  the  one-parameter 
 subgroup associated with the vector field $\cX$ on $\bM$ corresponding to an element of the Lie algebra
 of $PSL(2,\bR)$.
  In particular the vector field on $\bM$,  $\cD$,  
  generating a one-parameter subgroup of $PSL(2,\bR)$, 
 is that defined in in Eq. (16) in \cite{MP6}.
 If $\zeta\in L^2(\Sigma,\omega_\Sigma)$ and  $\lambda_\zeta$  is the 
  coherent state  
on $\cW(\bM)$
defined by means of (33) of Section IV B in \cite{MP6}, we indicate its  GNS triple 
by $(\gH_\zeta, \Pi_\zeta, \Psi_\zeta)$.
 States $\lambda_\zeta$ are constructed as follows with respect to $\lambda:= \lambda_0$.
   The map $V(\omega) \mapsto V(\omega) \:e^{i \int_\bM  \Gamma  (\zeta\omega_+ + \overline{\zeta\omega_+})}$, $\omega\in \cD(\bM)$, uniquely
extends to a $^*$-automorphism  $\gamma_\zeta$ on  $\cW(\bM)$ such that $\gamma_\zeta \circ \alpha_t^{(\cD)} = 
\alpha_t^{(\cD)} \circ \gamma_\zeta$ for all $t\in \bR$. 
The function $\Gamma:= \ln |\tan(\theta/2)|$ and the $\cD$-positive-frequency part of $\omega$, $\omega_+$, are respectively 
defined and discussed in section
 IV B (Eq. (32)) and in Lemma 3.1 of section III B of
\cite{MP6}.
\beq
\lambda_\zeta(w) &:=& \lambda(\gamma_\zeta w) \:,\:\:\:\:\: \mbox{for all $w\in \cW(\bM)$}\label{1}\:.
\eeq
The state $\lambda$ was defined in Eq. (13) in \cite{MP6} and it is pure in accordance with
 Lemma A.2 in \cite{KW},  because the real linear map $K: \cS(\bM) \ni \psi \mapsto \psi_+$
 has dense range in the one-particle space $\cH$ by construction.
It is clear that the GNS representation generated by $\lambda_\zeta$ is irreducible 
 if that of  $\lambda$ is irreducible. As a consequence every $\lambda_\zeta$ is pure.\\
 Among other results it has been proved that (theorems 4.1, 4.2, 5.1)  the  pure states 
 $\lambda_\zeta$  are inequivalent states and,  
 the restrictions to
 the algebra localized at the ``half circle times $\Sigma$'',
 give rise to different extremal KMS states
 at rationalized Hawking temperature $T=1/2\pi$ 
 with respect to $\{\alpha^{(\cD)}_t\}_{t\in \bR}$. If one is dealing with a bifurcate Killing Horizon of a black hole 
 the ``half circle times $\Sigma$'' is noting but $\bF_+$, 
the future right branch of the Killing  horizon (see Section I and figure in \cite{MP6}). In this case
 $-\zeta^{-1}\cD$, with $\zeta^{-1} = \kappa$ being the surface gravity of the examined black hole,
 can be recognized as the restriction to the horizon to the 
 Killing vector field defining Schwarzschild time, $kT$ becomes proper Hawking's temperature and 
 the order parameter associated with the breakdown of symmetry can be related with properties of the black hole.

\section{Spontaneous breaking of $PSL(2,\bR)$ symmetry.}
As illustrated  in further details below, a remarkable property of states $\lambda_\zeta$ is that, for $\zeta\neq 0$,
 they  break part of the symmetry, in the sense that each of them is not invariant under the full group $PSL(2,\bR)$ and so
there is no unitary representation of whole group $PSL(2,\bR)$ which leaves fixed the cyclic GNS vector.
We want here to clarify the nature of this breakdown of $PSL(2,\bR)$ symmetry, proving that, actually
there is no unitary representation of $PSL(2,\bR)$ which implements the full group action of $PSL(2,\bR)$
no matter the issue of the invariance of $\lambda_\zeta$ under $PSL(2,\bR)$. This leads to {\em spontaneous} breaking of $PSL(2,\bR)$ symmetry.\\  
 Let us focus on this issue from a general point of view. In the physical literature there are several, also strongly inequivalent, definitions 
 of {\em spontaneous} breaking of symmetry related to different approaches to quantum theories. 
 We  adopt the following elementary definition which is quite natural in algebraic QFT and  is equivalent 
 to the definition given on p.119 of the recent review \cite{Strocchi} on breaking symmetry theory (see the end of the paper
 for further comments). \\
 
 \noindent {\bf Definition 1}. {\em Referring to a $C^*$-algebra $\cA$, one says that:\\
 {\bf (a)} A Lie group $\cG$
 is a {\bf group of symmetries} for $\cA$ if there is a 
 representation $\beta$ of $\cG$ made of $*$-automorphisms of $\cA$.
 If some notion of time evolution is provided, it is required that it corresponds to a 
 one parameter subgroup of $\cG$\footnote{If this subgroup does not belong to the center of $\cG$, 
 the self-adjoint generators of unitary representations of $\cG$ implementing $\beta$, if any, give rise to constants of
 motion which may depend parametrically on time. It happens, for instance, for Poincar\'e-invariant systems, where
 the conserved self-adjoint operator  $K^i(t)$ associated with
boost invariance along $i$-th axis depend parametrically on time. Indeed, on the appropriate domain,
$K^i(t) = K^i(0) -t P^i$ and it is conserved under Heisenberg evolution
$U_t K^i(t)U^\dagger_t = K^i(0)$.}.  \\
 {\bf (b)}  Assuming that (a) is valid, {\bf spontaneous breaking of $\cG$ symmetry} occurs 
 with respect to an algebraic state $\mu$ on $\cA$ 
 and $\beta$,
 if  there are elements $g\in \cG$ such that $\beta_g$
 are not implementable unitarily  in the GNS representation $(\gH_\mu,\Pi_\mu,\Psi_\mu)$
 of $\mu$, i.e. for those elements there is no unitary operator $U_g$ on $\gH_\omega$ 
  with $U_g\Pi_\mu(a)U_g^\dagger = \Pi_\mu(\beta_g(a))$
 for all $a\in \cA$.}\\
 
 \noindent   Considering the inequivalent GNS triples $(\gH_\zeta, \Pi_\zeta, \Psi_\zeta)$, theorems 3.2
  and 3.3 in \cite{MP6} show that, if 
 $\zeta=0$ the group of automorphisms $\alpha$ representing $PSL(2,\bR)$ can be unitarily implemented 
 in the space $\gH=\gH_0$ and the cyclic vector $\Psi := \Psi_0$ of the GNS representation is invariant 
 under that (strongly continuous) unitary representation of $PSL(2,\bR)$\footnote{In the proof
  theorem 3.2
 in all occurrences of the symbol $PSL(2,\bR)$
 before the statement ``...is in fact a representation of $PSL(2,\bR)$.''
  it has to be replaced by $\widetilde{PSL(2,\bR)}$, it denoting the universal covering of $PSL(2,\bR)$.}.
 Conversely, if $\zeta\neq 0$, $PSL(2,\bR)$ symmetry turns out to be broken. Indeed,  theorem 4.1 states that 
 each state $\lambda_\zeta$ with $\zeta \neq 0$ is invariant under 
$\{\alpha^{(\cD)}_t\}_{t\in \bR}$, but  it is  not under any
other one-parameter subgroup of $\alpha$ (barring those associated with $c\cD$ for $c\in \bR$ constant). 
In the general case this is not enough to assure occurrence of {\em spontaneous}
 breaking of $PSL(2,\bR)$ symmetry as defined in Def.1. \\
In a different context, it is possible to show that -- see section III.3.2 of \cite{Haag} -- 
 if $\cG$ is Poincar\'e group or an internal symmetry group for a special relativistic 
 system and the reference state $\mu$ is a primary vacuum state over the  net of  algebras of observables, 
 then non-$\cG$ invariance of the vacuum state implies -- and in fact is equivalent to -- spontaneous breaking of $\cG$ symmetry.
 Our considered case is far from that extent and  thus there is no {\em a priori} guarantee for the occurrence   
of spontaneous breaking of $PSL(2,\bR)$ symmetry for 
 states $\lambda_\zeta$ with $\zeta\neq 0$ and the issue deserve further investigation.
 The following theorem give an answer to the issue.\\

\noindent {\bf Theorem 1}. {\em If $L^2(\Sigma, \omega_\Sigma) \ni \zeta \neq 0$, 
spontaneous breaking of $PSL(2,\bR)$-symmetry occurs
with respect to $\lambda_\zeta$ and the representation $\alpha$. 
(In particular, there is no unitary implementation of the nontrivial elements of the  subgroup
of $PSL(2,\bR)$ generated from the vector field $\frac{\partial}{\partial\theta}$.)}\\

\noindent {\em Proof}. 
 Referring to the GNS triple of $\lambda_\zeta$ define
$\hat{V}_\zeta(\omega) := \Pi_\zeta(V(\omega))$. 
The existence of a unitary implementation, $L_g: \gH_\zeta\to \gH_\zeta$,
of $\alpha$ in the GNS triple $(\gH_\zeta, \Pi_\zeta, \Psi_\zeta)$
implies, in particular,  that
\beq L_g \hat{V}_\zeta(\omega) L^\dagger_g = \Pi_\zeta\left(\alpha_g (V_\zeta(\omega))\right)\:, \quad \mbox{for all $\omega \in 
\cD(\bM)$ and every $g\in PSL(2,\bR)$}\:. \label{mission:impossible}\eeq
 By construction, $(\gH_\zeta, \Pi, \Psi_\zeta)$ (notice that we wrote $\Pi$ instead of $\Pi_\zeta$) 
is a GNS triple of $\lambda$ (notice that we wrote $\lambda$ instead of $\lambda_\zeta$)  if 
 \beq
 \Pi : V(\omega) \mapsto  \hat{V}(\omega) := \hat{V}_\zeta(\omega) e^{-i \int_\bM  \Gamma  (\zeta\omega_+ + 
 \overline{\zeta\omega_+})}\label{fine}\eeq
 In this realization $\alpha$ can be unitarily implemented 
(theorem 3.2 in \cite{MP6}): There is a (strongly continuous) unitary representation $U$ of $PSL(2,\bR)$ 
such that
\beq U_g  \hat{V}(\omega)U^\dagger_g = \Pi\left(\alpha_g({V}(\omega))\right)\:,\quad \mbox{for all $\omega \in 
\cD(\bM)$ and every $g\in PSL(2,\bR)$}\:.\label{A}\eeq
Suppose  that $\alpha$ can be implemented also in $(\gH_\zeta, \Pi_\zeta, \Psi_\zeta)$, where now
$\Pi_\zeta : V(\omega) \mapsto  \hat{V}_\zeta(\omega)$,
 and let $L$ be the corresponding unitary representation of $PSL(2,\bR)$ satisfying (\ref{mission:impossible}).
That equation together with (\ref{fine}) entail that the unitary operator $S_g := U_g^\dagger L_g$ satisfies
\beq S_g V(\omega)S_g^\dagger &=& e^{ic_{g,\omega}}V (\omega)\:,\label{PG}\\
c_{g,\omega} &:=& \int_\bM    
\left(\zeta\left(\omega^{(g^{-1})}\right)_+ + \overline{\zeta\left(\omega^{(g^{-1})}\right)_+} 
-  \zeta\omega_+ - \overline{\zeta\omega_+}\right) \Gamma\:. \nonumber \eeq
Now, dealing with exactly as in the proof of (ii) of (b) of theorem 4.1 (where the role 
 of our $S_g$ was played by the operator $U$ and the role of  $c_{g,\omega}$  
 was played by the simpler phase $\int_\bM  (\zeta\omega_+ + \overline{\zeta\omega_+}) \Gamma $)
 one finds that (\ref{PG}) entails that $\langle \Psi_\zeta, S_g \Psi_\zeta \rangle \neq 0$ and 
 \beq ||S_g\Psi_\zeta||^2 = 
 |\langle \Psi_\zeta, S_g \Psi_\zeta \rangle |^2 \:\: e^{ \sum_{n,j}^\infty |\eta_{n,j}|^2}\:,\label{div}\\
 \eta_{n,j} := -2i \int_\bM \Gamma(\theta)\overline{\zeta(s)} u_j(s) \:\frac{\partial}{\partial \theta}
\frac{e^{i\theta_g} - e^{i\theta}}{\sqrt{4\pi n}}\: d\theta d\omega_\Sigma(s)\:.\nonumber \eeq
Above the real compactly-supported functions $u_j$ defines a Hilbert base in $L^2(\Sigma, \omega_\Sigma)$,
 and $(\theta_g,s_g) := g(\theta,s)\in \bM$ is obtained by the action of $g\in PSL(2,\bR)$ on $(\theta,s)\in \bM$
 (obviously $s_g=s$ since $PSL(2,\bR)$ acts on the factor $\bS^1$ of $\bM = \bS^1\times \Sigma$). 
Now take $g\in\{\alpha^{(\cK)}_t\}_{t\in \bR}$, the one-parameter subgroup of $PSL(2,\bR)$
generated by the vector field
$\cK := \frac{\partial}{\partial\theta}$, and realize the factor $\bS^1$ of $\bM=\bS^1\times \Sigma$ as $[-\pi,\pi)$ with the
identification of its  endpoints.
In this case, obviously, $\theta_g = \theta +t$.
A direct computation shows that, for some $j_0$ with $\int_\Sigma u_j(s)\overline{\zeta(s)} \omega_\Sigma(s) =0$ (which does exist otherwise
$\zeta =0$ almost everywhere)
$$|\eta_{2n+1,j_0}|^2 = C \left[ \frac{1}{2n+1} - \frac{\cos((2n+1)t)}{2n+1} \right]$$
 for some constant $C>0$ independent form $n$. The series of elements $-\cos((2n+1)t)/(2n+1)$ converges for $t\neq 0,\pm\pi$ (it diverges to $+\infty$
 for $t=\pm \pi$), whereas that of elements $1/(2n+1)$
 diverges to $+\infty$. Thus the exponent in (\ref{div}) and 
 $L_g$ cannot exist, if $g= \alpha^{(\cK)}_t$ with $t\neq 0$. $\Box$\\

\noindent{\bf Remark}. The automorphism $\gamma_\zeta$  is a symmetry of 
  the system because it commutes with time evolution
$\alpha_t$. (It is worth noticing that, if restricting to real functions $\zeta$, 
$\zeta \mapsto \gamma_\zeta$ defines a group of automorphisms.)
 Since 
 $\lambda_\zeta\spa\rest_{\cW(\bF_+)}\neq \lambda_{\zeta'}\spa\rest_{\cW(\bF_+)}$ for $\zeta\neq \zeta'$,
and  all these states are extremal  $\alpha^{(\cD)}$-KMS states 
at the same temperature, following Haag (V.I.5 in \cite{Haag}), 
we can say that {\em spontaneous symmetry breaking} with respect to $\gamma_\zeta$ occurs in the
context of extremal KMS states theory.

 \section{Final comments on inequivalent approaches concerning symmetry breakdown}  
If $\mu$ is a state on the  $C^*$-algebra $\cA$, 
its {\em Gelfand ideal}, $\cI_\mu$, consists of the elements $a\in \cA$ such that $\mu(a^*a)=0$.  $\cI_\mu$ plays a central role 
in GNS reconstruction procedure. Breakdown of symmetry could be investigated 
from another point of view -- relying upon the invariance properties of Gelfand ideal -- with some overlap with our approach. 
It is based on the following proposition.\\

\noindent {\bf Proposition 1}. 
{\em Let $\beta$ be a faithful $*$-automorphism representation of the group $\cG$ on the $C^*$-algebra $\cA$ with unit $\bI$ and let 
$\mu: \cA \to \bC$ be a 
state
with GNS triple $(\gH_\mu, \Pi_\mu, \Psi_\mu)$.\\
The Gelfand ideal $\cI_\mu$ is invariant under $\beta$ if and only if there is a $\cG$-representation made 
 of densely-defined operators $U_g: \Pi_\mu(\cA) \to \Pi_\mu(\cA)$ which implements $\beta$ 
leaving fixed $\Psi_\mu$, i.e.
\beq (i) \:\:\:\: U_g \Pi_\mu(a) U^{-1}_g = \Pi_\mu(\beta_g (a))\: \:\mbox{for all $a\in \cA, g\in \cG\:\:$
and  $(ii) \:\:\:\: U_g \Psi_\mu = \Psi_\mu$ for all $g\in \cG$\:.} \label{fine2}\eeq
If a representation  $\cG\ni g \mapsto U_g$ satisfying (\ref{fine2}) exists the following holds.\\
{\bf (a)} It is unique and the operators $U_g$ are completely determined 
by  \beq U_g \Pi_\mu(a) \Psi_\mu = \Pi_\mu(\beta_g a)\Psi_\mu\:, \quad \mbox{for all $a\in \cA$ and $g\in \cG$.}\label{fine3}\eeq
{\bf (b)} Operators $U_g$ are (restrictions to the dense domain $\Pi_\mu(\cA)$ of uniquely determined) unitary operators on $\gH_\mu$,
if and only if $\mu$ is invariant under $\beta$.}\\

\noindent{\em Sketch of proof}. From GNS theorem  $\cI_\mu = \{a\in \cA\:|\: \Pi_\mu(a)\Psi_\mu = 0\}$.
From it one easily proves that if $\cI_\mu$ is $\beta$-invariant (\ref{fine3}) 
define a well-posed representation of $\cG$ satisfying (\ref{fine2}) (notice that,
$\beta_g(\bI)=\bI$ and $\Pi_\mu(\bI)=I$ so that (ii) in (\ref{fine2}) holds true from (\ref{fine3})).
Since (i) and (ii) in (\ref{fine2})
entails (\ref{fine3}), $\cI_\mu=\{a\in \cA\:|\: \Pi_\mu(a)\Psi_\mu = 0\}$ proves that the existence 
of a representation satisfying (\ref{fine2}) implies $\beta$-invariance of $\cI_\mu$ and (a) is valid as well. 
Proofs of (b) is based on the identity (from  invariance of $\mu$ and GNS theorem) $||U_g\Pi_\mu(a) \Psi_\mu||^2 = \mu(\beta_g(a^*a)) = \mu(a^*a) = 
||\Pi_\mu(a) \Psi_\mu||^2$. $\Box$\\

\noindent In view of that result, by a pure mathematical point of view, a strategy to distinguish several degrees of $\cG$-symmetry
 breakdowns for a state $\mu: \cA \to \bC$ when $\cG$ is represented, at algebraic level, by the $*$-automorphism representation
 $\beta$, could be the following.

(1) ({\em No symmetry breaking}) Gelfand ideal is invariant under $\beta_g$  -- i.e. (i) and (ii) in (\ref{fine2}) 
are valid for an operator
representation of $\cG$. 

(2)  Gelfand ideal is invariant under $\beta$ -- so that both (i) and (ii) in (\ref{fine2})
hold true -- but the induced action of $\beta$ in the GNS representation of $\mu$ is {\em not} unitarily implementable.

(3) Gelfand ideal is {\em not} invariant under $\beta_g$ -- so that at least one of (i) and (ii) in (\ref{fine2}) is not valid
for any operator representation of $\cG$ on the relevant domain.

\noindent In this paper we, instead,  have adopted quite a different point of view (sharing however some overlap
with the point of view illustrated above) entirely  based on the definition of spontaneous breaking of symmetry 
defined on p. 119 of \cite{Strocchi}. We considered only the problem of {\em unitary} (non)implementability of ($PSL(2,\bR)$) 
symmetry. This is in accordance with the well-known general Wigner-Kadison notion of {\em quantum symmetry} described in terms of appropriate 
(projective) unitary or anti-unitary operators \cite{Wsymmetries}. 
The three degree of symmetry breakdown one may consider in this context are the following, referring to $\mu,\cG,\beta$ as before. 

(U1) ({\em No symmetry breaking}) Algebraic symmetry $\beta$ is implementable for $\mu$ by means of unitary operators
{\em and} the state $\mu$ is invariant under the  action $\beta$ of the full group $\cG$ -- (i) and (ii) in (\ref{fine2}) are valid for a unitary
representation of $\cG$. 

(U2) ({\em Symmetry breaking due to the cyclic vector}) Algebraic symmetry $\beta$ is implementable for $\mu$ by means of 
unitary operators
{\em and} the state $\mu$ is {\em not} invariant under the action $\beta$ of the full group $\cG$ -- (i), but not  (ii) in (\ref{fine2}), is
 valid for a unitary representation of $\cG$. 

(U3) ({\em Spontaneous symmetry breaking}) Algebraic symmetry $\beta$ is not 
implementable for $\mu$ by means of unitary operators -- (i) in (\ref{fine2}) does not hold for any unitary representation of $\cG$.

\noindent In \cite{MP6} we established, for states $\lambda_\zeta$ with $\zeta\neq 0$ the validity of either degree (U2)
 or (U3) concerning 
$\cG = PSL(2,\bR)$ symmetry breaking. This paper shows that, actually, the strongest degree (U3) takes place.\\
 {\em A priori} this extent may be compatible 
with either (2) or (3) of the other scheme. A closed scrutiny would be necessary to examine this issue but it is far from the goal of this paper
where only Wigner-Kadison unitary symmetries are considered since they are the only which preserve quantum probabilities and 
have direct physical meaning.\\
 To conclude we notice that  an example of the intermediate case 
(U2) is realized  in the framework of \cite{MP6} with respect to the Weyl algebra $\cW(\bM)$, with $\bM = \bS^1$, and for
the fully $PSL(2,\bR)$-symmetric state $\lambda_{\zeta=0}$.
Breaking of symmetry at level (U2) occurs when extending the symmetry group  from $PSL(2,\bR)$ to the 
infinite dimensional one $\cG := Diff^+(\bS^1)$.



\begin{thebibliography}{999}

\bibitem{MP6} V.~Moretti  and N.~Pinamonti: 
{\em ``Bose-Einstein condensate  and Spontaneous Breaking of  Conformal Symmetry on Killing Horizons''},
J. Math. Phys. 46, 062303 (2005). [hep-th/0407256]


\bibitem{KW} B.~S.~Kay, R.~M.~Wald,
{\em ``Theorems On The Uniqueness And Thermal Properties Of Stationary, Nonsingular,
Quasifree States On Space-Times With A Bifurcate Killing Horizon'',}
Phys.\ Rept.\  {\bf 207}, 49 (1991).


\bibitem{Haag} R.~Haag,
{\em ``Local quantum physics: Fields, particles, algebras''},
Second Revised and Enlarged Edition.
Springer Berlin, Germany  (1992).  


\bibitem{Strocchi} F.~Strocchi,
{\em ``Symmetry Breaking''},
Lecture Notes in Physics,  
Springer Berlin, Germany  (2005).  


\bibitem{Wsymmetries} R.~Kadison, {\em ``Isometries of Operator Algebras''},  Ann. Of Math. (2)  {\bf 54}, 325  (1951),\\ 
C.~Piron, {\em ``Foundations of Quantum Physics''}, W. A. Benjamin, London
(1976).


\end{thebibliography}
\end{document}